\documentclass[aps,twocolumn,superscriptaddress,PRD]{revtex4}
\usepackage{amsmath,bm}

\usepackage{amsmath}
\usepackage{epsfig}
\usepackage{multirow}
\usepackage{slashed}
\usepackage{amsmath}
\usepackage{color}
\usepackage{float}
\usepackage{graphicx}
\usepackage[colorlinks,
            linkcolor=blue,
            anchorcolor=blue,
            citecolor=blue
            ]{hyperref}

\usepackage{diagbox}
\usepackage{booktabs}
\usepackage{subfigure}
\graphicspath{{figures/}} 
\def\bea#1\eea{\begin{align}#1\end{align}}

\newcommand{\bef}{\begin{figure}[!htp]}
\newcommand{\eef}{\end{figure}}

\begin{document}
\title{Suppression of elliptic anisotropy inside jets: A new perspective for jet quenching}

\author{Meng-Quan Yang}
\affiliation{Key Laboratory of Quark $\&$ Lepton Physics (MOE) and Institute of Particle Physics, Central China Normal University, Wuhan 430079, China}

\author{Wei-Xi Kong}
\affiliation{Key Laboratory of Quark $\&$ Lepton Physics (MOE) and Institute of Particle Physics, Central China Normal University, Wuhan 430079, China}

\author{Peng Ru}
\email{p.ru@m.scnu.edu.cn}
\affiliation{School of Materials and New Energy, South China Normal University, Shanwei 516699, China}

\affiliation{Guangdong-Hong Kong Joint Laboratory of Quantum Matter,
Guangdong Provincial Key Laboratory of Nuclear Science, Southern Nuclear Science Computing Center, South China Normal University, Guangzhou 510006, China}

\author{Ben-Wei Zhang}
\email{bwzhang@mail.ccnu.edu.cn}
\affiliation{Key Laboratory of Quark $\&$ Lepton Physics (MOE) and Institute of Particle Physics, Central China Normal University, Wuhan 430079, China}

\date{\today}

\begin{abstract}
Particle azimuthal anisotropies inside jets, defined within the momentum plane perpendicular to the jet axis, carry the information of the QCD cascade process for jet formation. In this work, we propose to measure the medium-induced modifications of the elliptic anisotropy inside jets in relativistic heavy-ion collisions to provide novel insight into the jet quenching phenomenon. By simulating the jet propagation in the hot and dense nuclear medium with a Linear Boltzmann Transport model, we observe a de-correlation in the two-particle azimuthal angular distribution for inclusive jet production in AA collisions relative to that in $pp$ collisions, which results in significant suppression of the in-jet elliptic anisotropy coefficient $v_2$. This phenomenon arises from the stochastic and strong interactions with the thermal QGP medium undergone by the jet particles. Furthermore, the nuclear modifications of the in-jet $v_2$ are found to be sensitive to the medium properties in the model study, which provide a potential probe for the jet tomography of nuclear matter.
\end{abstract}


\maketitle

\section{introduction}
\label{sec:Intro}
Jet production has been long regarded as a powerful tool for exploring the nature and properties of Quantum Chromodynamics~(QCD) in high-energy collider physics~\cite{Sapeta:2015gee,Larkoski:2017jix,Gross:2022hyw}. In nuclear collisions, jets also act as excellent probes of the properties of both cold nuclei and hot quark-gluon plasma~(QGP), which reveal fundamental non-perturbative structures of these nuclear matters~\cite{Cao:2020wlm,Cunqueiro:2021wls,JET:2013cls,JETSCAPE:2021ehl,Budhraja:2024ttc,Andres:2022ovj,Yang:2023dwc,Ru:2019qvz,Barata:2024wsu,Fu:2024pic,Xie:2024xbn}.

As an ensemble of many-body systems, jets can resolve medium properties at multiple observation dimensions~\cite{Budhraja:2024ttc}. Earlier studies focused on low-dimensional observables in which a jet is treated as a whole object. For instance, the suppressed jet/hadron spectra, related to the parton energy loss in nuclear medium, have been a landmark of the jet quenching phenomena~\cite{STAR:2003fka,PHENIX:2003qdj,ATLAS:2010isq,ALICE:2013dpt,CMS:2012ytf,Gyulassy:2003mc,Qin:2015srf,Qin:2007rn,Majumder:2011uk,Li:2024uzk}.
In the past decade, with both the experimental and theoretical developments, a series of higher-dimensional observables, i.e., the jet substructures and their nuclear modifications, which provide more details on the jet-medium interactions, have attracted a lot of attentions. Typical jet substructures, like jet shape~\cite{CMS:2013lhm,Vitev:2008rz,Chang:2019sae,Kang:2017mda,KunnawalkamElayavalli:2017hxo} and fragmentation functions~\cite{CMS:2018mqn,Chen:2020tbl}, can be classified as single-inclusive distributions inside jets. With recent applications of the energy-energy correlators~\cite{Komiske:2022enw}, a type of double-inclusive distribution or angular correlation inside jets, remarkable progresses have been made on disclosing novel structures of the medium~\cite{Andres:2022ovj,Yang:2023dwc,Barata:2024wsu,Fu:2024pic,Chen:2024cgx,Shen:2024oif}.

On the other hand, azimuthal-angle correlations~\cite{Ollitrault:1992bk,Bilandzic:2010jr,Voloshin:2008dg} of the particles emitted from collision systems have been extensively used to infer the properties of the systems, such as thermalization and collectivity~\cite{PHENIX:2018lia,STAR:2024wgy,CMS:2010ifv,ALICE:2012eyl,LHCb:2015coe,ATLAS:2015hzw,CMS:2016fnw,ATLAS:2021jhn,Heinz:2013th,Gale:2013da,Zhao:2017rgg,Ru:2019deq,Zhao:2022ayk}. Recently, under the viewpoint that jets are a sort of small collision systems~\cite{Baty:2021ugw}, two-particle azimuthal-angle correlations have been successfully employed to study the possible emergence of an in-jet collectivity in hadronic collisions~\cite{CMS:2023iam,Zhao:2024wqs}. Since such correlations are sensitive to the system properties, how they will be modified in heavy-ion collisions and what medium properties can be decoded from the two-particle correlation inside jets, a type of double-inclusive observable, is of great interest.

In high-energy collision systems, azimuthal-angle correlations mainly arise in forms of azimuthal anisotropies in momentum space~\cite{Ollitrault:1992bk,Bilandzic:2010jr,Voloshin:2008dg,PHENIX:2018lia,STAR:2024wgy}. For jet systems, azimuthal angle is defined within the momentum plane perpendicular to the jet axis~\cite{Baty:2021ugw}, and azimuthal anisotropies can be naturally generated in the QCD cascade process of the jets in vacuum. Theoretical descriptions of the generation of the in-jet azimuthal-angle correlations, which involves non-perturbative dynamics, should be quite complicated. However, it is noteworthy that the current event generators can well describe the data of two-particle correlations in hadronic collisions, except for a very high multiplicity region~\cite{CMS:2023iam,Zhao:2024wqs}. Therefore, one can expect that the generators are able to capture the main features of the correlations, such as in an observation binned with the transverse momentum of the inclusive jet, where the rarely produced high-multiplicity jets give negligible contributions, and to provide a reasonable baseline to study the medium-induced modifications.

In this work, we study the nuclear modifications on the elliptic anisotropy inside jets in relativistic nucleus-nucleus~(AA) collisions, based on the inclusive jet events in proton-proton~($pp$) collisions generated by PYTHIA8~\cite{Sjostrand:2014zea,Bierlich:2022pfr}. By simulating the jet propagation in a hot and dense QGP medium with a Linear Boltzmann Transport~(LBT) model~\cite{Wang:2013cia,Luo:2018pto,Li:2010ts,He:2015pra,Cao:2016gvr,Cao:2017hhk}, we observe an obvious de-correlation in the two-particle azimuthal-angle distribution in AA collisions relative to that in $pp$ collisions, which results in significant suppression of the in-jet elliptic anisotropy coefficient $v_2$. Such an effect indicates that, as an irreversible thermodynamical process, the jet transport in QGP medium may create more entropy than that in vacuum, and the information of the earlier-time evolution of a jet is more likely lost in the medium. The modifications on $v_2$ inside jets are expected to provide a new perspective for understanding the jet quenching phenomena.

The remainder of this paper is organized as follows. In Sec.~\ref{sec:framework}, we discuss the azimuthal anisotropy inside jets in elementary $pp$ collisions and show the results of $v_2$ with the PHTHIA8 model. In Sec.~\ref{sec:results}, we shall study the nuclear modifications on the in-jet $v_2$ for inclusive jet production with the LBT model.
A summary and discussion is presented in Sec.~\ref{sec:summary}.

\section{Elliptic anisotropies inside jets in elementary hadronic collisions}
\label{sec:framework}
The azimuthal-angle correlations inside jets are defined within the jet coordinate frame in momentum space proposed in Refs.~\cite{Baty:2021ugw,CMS:2023iam}. The main idea is to set the longitudinal direction along the jet axis by considering each jet as a collision system. In this way, the momenta of the constituent particles of each jet can be expressed in the corresponding frame as $\vec{p^*}=(j_T, \eta^*, \phi^*)$, where $j_T$ and $\phi^*$ represent the magnitude and azimuthal angle of the transverse momentum $\vec{j}_T$, respectively, and $\eta^*$ denotes the pseudorapidity characterizing the angle between particle momentum and longitudinal direction. 

Since the longitudinal direction is parallel to the total momentum of a jet system, the vector sum over the transverse momenta of all the constituent particles should always vanish, expressed as $\sum_i{\vec{j}_{T,i}}=0$. Such a condition is an important source of azimuthal-angle correlations when the number of constituent particles is not large. In a simple case when there is only two particles in each of the jets, the two constituents will always be back to back in transverse plane, which naturally generates an azimuthal anisotropy. Taking no account of the other dynamics-induced correlations, this kind of momentum anisotropy will decrease with an increasing number of constituents in general.

The various dynamical processes inherent in jet evolution and formation constitute another source of azimuthal-angle correlations. For example, in the  parton shower stage, the parton splittings are more likely to happen at smaller angles~\cite{Sjostrand:2014zea}, which may enhances the probability to detect two particles near to each other in direction. In a typical configuration where there is only one large-angle splitting but much more small-angle splittings inside a jet, the jet may be roughly thought as two sub-jets. In this case, two particles from a same sub-jet will contribute to the correlations at a short-range~(small differences in $\eta^*$), whereas those from different sub-jets may give rise to a back-to-back correlation, probably at a long-range. Besides the parton shower dynamics, the non-perturbative dynamics for the hadronization or fragmentation process may also affect the correlations to some extent. In addition, the interactions among the jet particles, at both partonic and hadronic stages, may have impacts on the azimuthal anisotropies. In particular, for the high-multiplicity jet production, in which the anisotropies arise from other sources become diminished, the rescatterings among the jet particles may lead to a hydrodynamics-like collective evolution, which may generate a long-range elliptic flow correlation~\cite{CMS:2023iam,Zhao:2024wqs}. 
 \begin{figure}[t]
 \vspace{-15pt}
\includegraphics[width=0.5\textwidth]{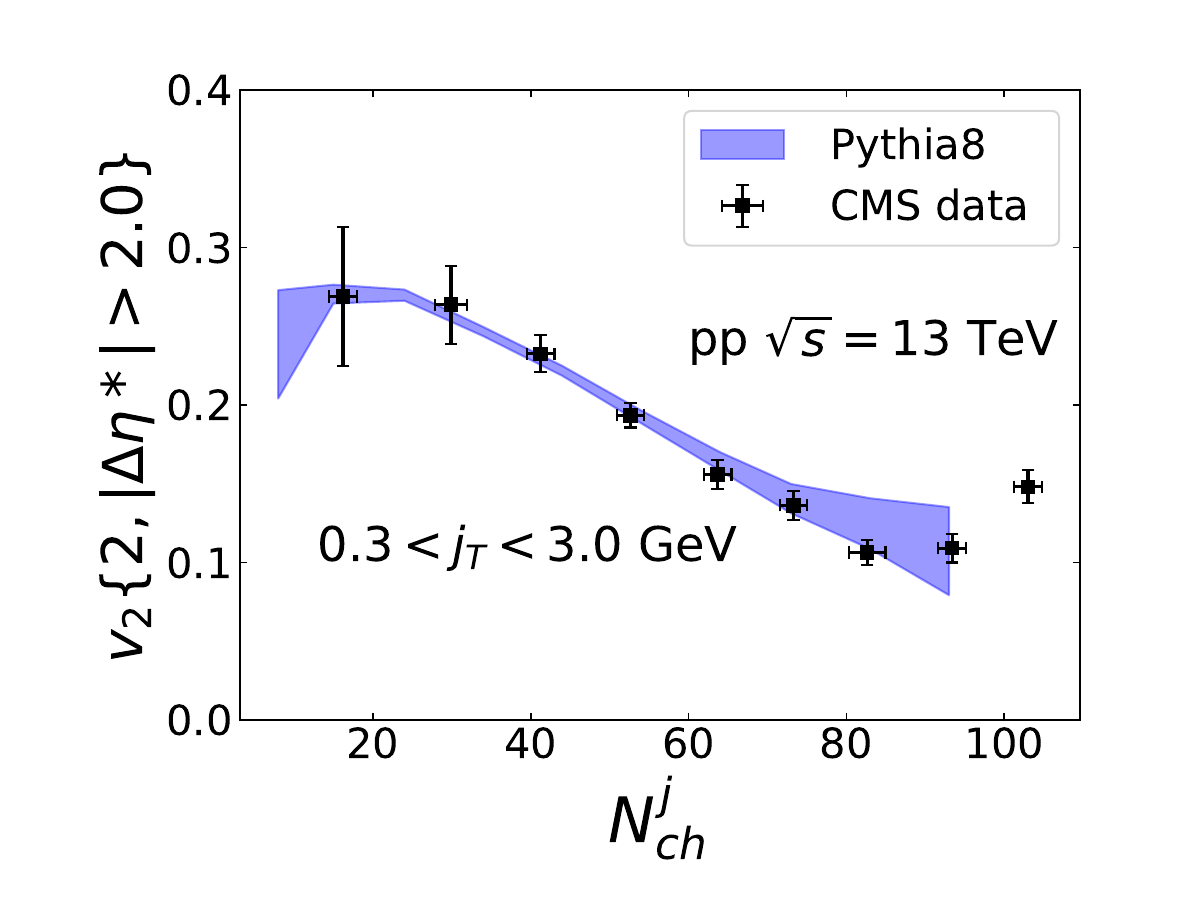}
  \caption{(Color online) Elliptic anisotropies coefficient $v_2\{2,|\Delta \eta^*|>2.0\}$ as a function of charged-particle multiplicity $N^j_{ch}$ inside jets in $pp$ collisions at $\sqrt{s}=13$~TeV. Black disks represent CMS data \cite{CMS:2023iam}.}\label{fig:CMS}
\end{figure}

In general, the particle azimuthal anisotropies inside jets carry the information of the QCD cascade process for jet formation, and all the dynamics-induced correlations can be modified with the presence of nuclear medium. To have a reasonable baseline for the study of medium modifications, we employ a Monte-Carlo event generator PYTHIA8~\cite{Sjostrand:2014zea,Bierlich:2022pfr} with the CP5 tune~\cite{CMS:2019csb} to simulate the events of jet production in elementary $pp$ collisions.

Figure~\ref{fig:CMS} shows a comparison between the model results and the CMS data~\cite{CMS:2023iam} for the elliptic anisotropy coefficient $v_2$ inside jets in $pp$ collisions as a function of the charged-particle multiplicity inside jets. Here the anti-$k_T$ jets with $R=0.8$~\cite{Cacciari:2011ma} are selected in the kinematic window $p_{T}>550$~GeV and $|\eta|<1.6$, and the $v_2\{2\}$ from two-particle correlations with a pseudorapidity gap $|\Delta\eta^*|>2$ are obtained using the charged particles inside jets with the cuts $|\eta|<2.4$, $p_{T}>0.3$~GeV and $0.3<j_T<3.0$~GeV~\cite{CMS:2023iam}. A Q-cumulant method~\cite{Bilandzic:2010jr, Zhao:2024wqs} is used to calculate the $v_2$ in the model predictions. 
\begin{figure*}[htbp]
    \centering
    \subfigure{
        \includegraphics[width=0.45\textwidth]{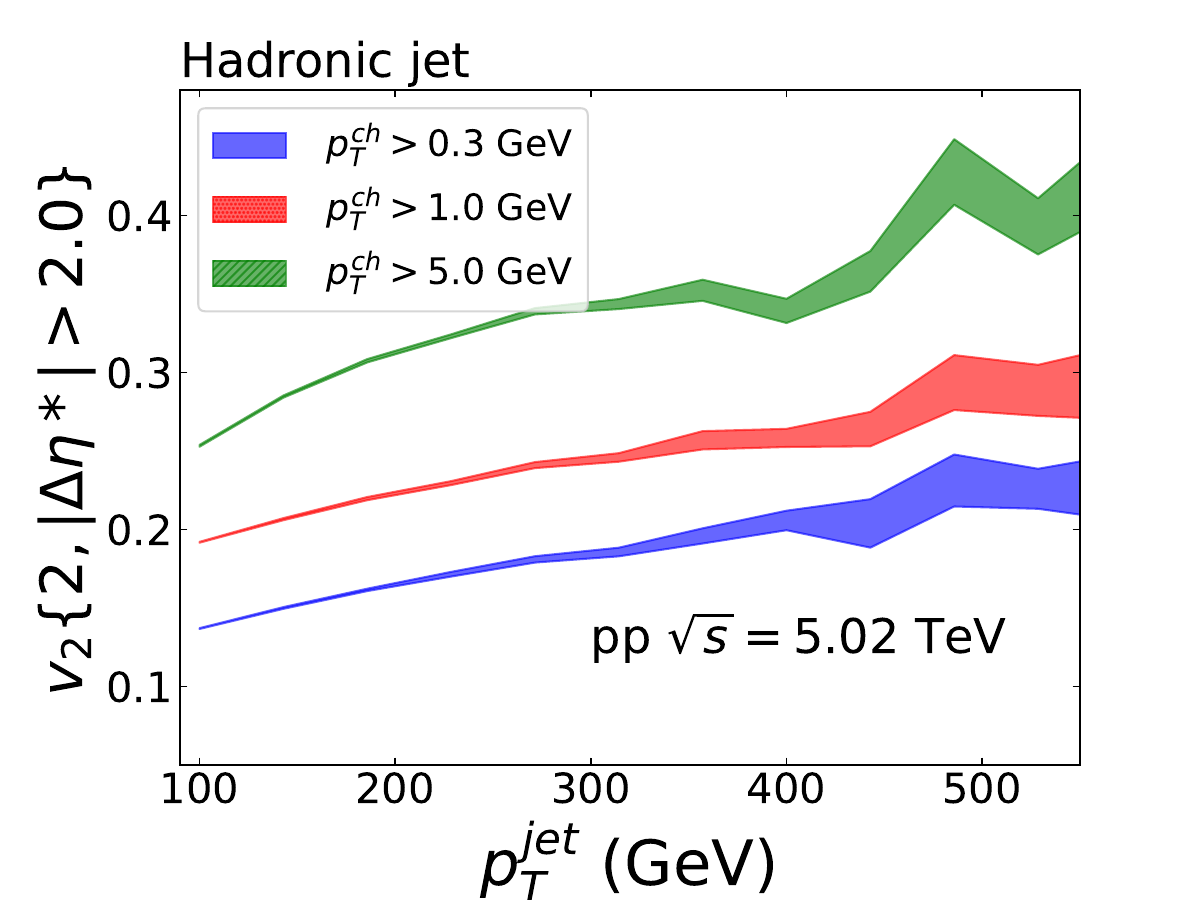}
    }
    \hspace{0.0\textwidth} 
    \subfigure{
        \includegraphics[width=0.45\textwidth]{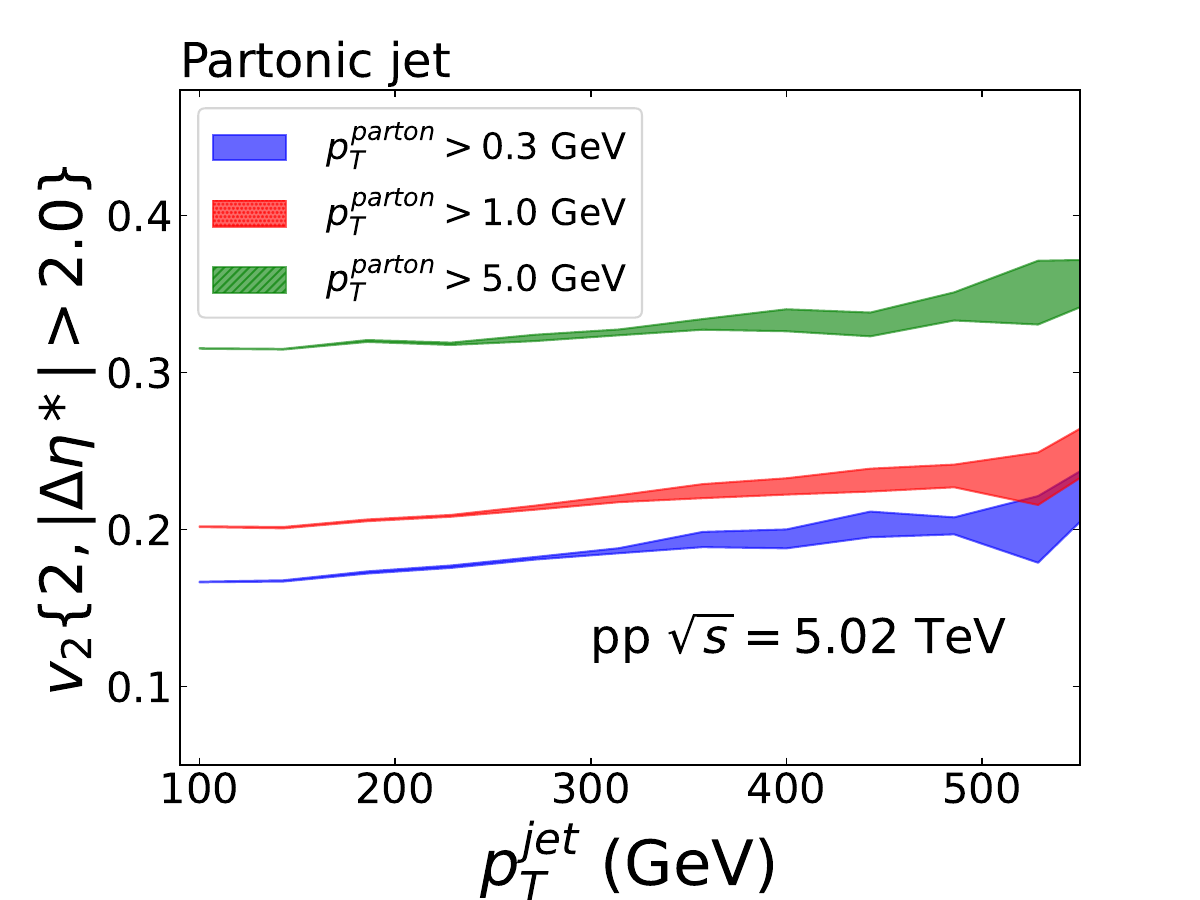}
    }
    \caption{(Color online) Elliptic anisotropies coefficient $v_2\{2,|\Delta \eta^*|>2.0\}$ as a function of jet transverse momentum $p_T^{\textrm{jet}}$ for inclusive jet production in $pp$ collisions at $\sqrt{s}=5.02$~TeV. 
    Results for jets reconstructed at both hadronic~(left panel) and partonic~(right panel) levels calculated with three cuts on transverse momenta of jet particles, i.e. $p_T>0.3$,  $p_T>1.0$ and  $p_T>5.0$ GeV are shown respectively. Anti-$k_T$ jets with $R=0.8$ selected in region $|\eta|<1.6$ are used in calculations.}
    \label{fig:v2_pp_inclusive}
\end{figure*}

As shown in Fig.~\ref{fig:CMS}, in most of the multiplicity regions where a downtrend of $v_2$ is observed, the CMS data can be well reproduced by the model calculations, which is in consistent with the results in Refs.~\cite{CMS:2023iam,Zhao:2024wqs}. In the very high multiplicity region, i.e. $N^j_{ch}>80$, the data begin to exhibit an uptrend. This feature was shown to be a possible signal of the collectivity arising from the rescatterings of the jet particles~\cite{CMS:2023iam,Zhao:2024wqs}, which have not been taken into account in this model calculations.
On the other hand, since this work focuses on the nuclear modifications for the inclusive jet production in which high-multiplicity jets have negligible contributions, the calculations are performed with a not very high statistical accuracy for the high-multiplicity cases. Besides the results of $v_2$, we also calculate the $\eta^*$ distribution of the charged hadrons inside jets, and find the results are in agreement with those in Refs.~\cite{CMS:2023iam,Zhao:2024wqs}. 

Given PYTHIA8's demonstrated ability to reasonably describe the $v_2$ data across most multiplicity regimes~(excluding high multiplicity regions), we perform a calculation for the inclusive jet production in $pp$ collisions at $\sqrt{s}=5.02$~TeV and plot the results of the $v_2$ inside jets as a function of the transverse momentum of jet in Fig.~\ref{fig:v2_pp_inclusive}. The left panel shows the results of $v_2\{2,|\Delta \eta^*|>2.0\}$ for the charged particles inside jets with three cuts on the particle transverse momentum, i.e. $p_T^{ch}>0.3$, $1.0$ and $5.0$~GeV, respectively. It is observed that in the studied regions the $v_2$ values are nontrivial, which slightly increase with the $p_T^{jet}$ for a certain $p_T^{ch}$ cut. At the same time, the $v_2$ obviously increase with an increasing $p_T^{ch}$ cut, which is partly due to that the harder particles which carry larger fractions of the jet momenta are more strongly limited by the momentum balance condition $\sum_i{\vec{j}_{T,i}}=0$. 

To study the role played by the parton shower dynamics in the generation of the $v_2$ inside jets, we plot in the right panel of Fig.~\ref{fig:v2_pp_inclusive} the $v_2\{2,|\Delta \eta^*|>2.0\}$ calculated with the jets reconstructed at the patonic level in PYTHIA8. Similar results as those for the hadronic jets can be observed, indicating that the hadronization with the Lund string fragmentation in PYTHIA8~\cite{Sjostrand:2014zea,Bierlich:2022pfr} may not bring a radical change in the pattern of the elliptic anisotropy in the inclusive jets. 

\section{Medium-induced modifications for inclusive jets in AA collisions}
\label{sec:results}
To study the medium-induced modifications on the elliptic anisotropy inside jets for inclusive jet production in AA collisions, we use an LBT model~\cite{Wang:2013cia,Luo:2018pto,Li:2010ts,He:2015pra,Cao:2016gvr,Cao:2017hhk} to simulate the in-medium evolution of the partonic-level jet events generated by PYTHIA8. Transport processes of both the jet shower partons and the jet-induced medium recoiled partons are included in LBT model, with their elastic scatterings with thermal medium and medium-induced radiations~(inelastic scatterings)~\cite{Guo:2000nz,Wang:2001ifa,Zhang:2003yn,Zhang:2003wk,Zhang:2004qm} being tracked. 

\begin{table}[b]
    \centering
    \renewcommand{\arraystretch}{1.6}
    \label{tab:ph}
    \begin{tabular}{c|c}
       \hline
        \,\,T~(GeV) \,\,$|$\,\,\,Prop. time~(fm/c) \,\,$|$\,\,\,\,$\alpha_s$ \,\,\,$|$\,\,\,\,\,\,Scatt.\,\,  & \,\,\,$\langle v_2\rangle$\,\,\,\\\hline 
       \textbf{A}:\,\,\,\,\,\,0.4, \,\,\,\,\,1.0, \,\,\,\,\,0.15, \,\,\,\,\,\,el.+inel.& 0.126\\\hline
       \textbf{B}:\,\,\,\,\,\,0.8, \,\,\,\,\,1.0, \,\,\,\,\,0.15, \,\,\,\,\,\,el.+inel.& 0.096\\\hline
       \textbf{C}:\,\,\,\,\,\,0.4, \,\,\,\,\,5.0, \,\,\,\,\,0.15, \,\,\,\,\,\,el.+inel.& 0.039\\\hline
       \textbf{D}:\,\,\,\,\,\,0.4, \,\,\,\,\,1.0, \,\,\,\,\,0.30, \,\,\,\,\,\,el.+inel.& 0.092\\\hline
       \textbf{E}:\,\,\,\,\,\,0.4, \,\,\,\,\,1.0, \,\,\,\,\,0.15, \,\,\,\,\,\,el. only& 0.147\\\hline\hline
       No medium~($pp$)  & 0.169  \\\hline 
    \end{tabular}
    \caption{$v_2\{2,|\Delta \eta^*|>2.0\}$ inside jets for inclusive jet production after propagating through a uniform static medium with different parameters for medium.}
    \label{tab:static}
\end{table}

 \begin{figure}[t]
 \vspace{-15pt}
  \centering
\includegraphics[width=0.45\textwidth]{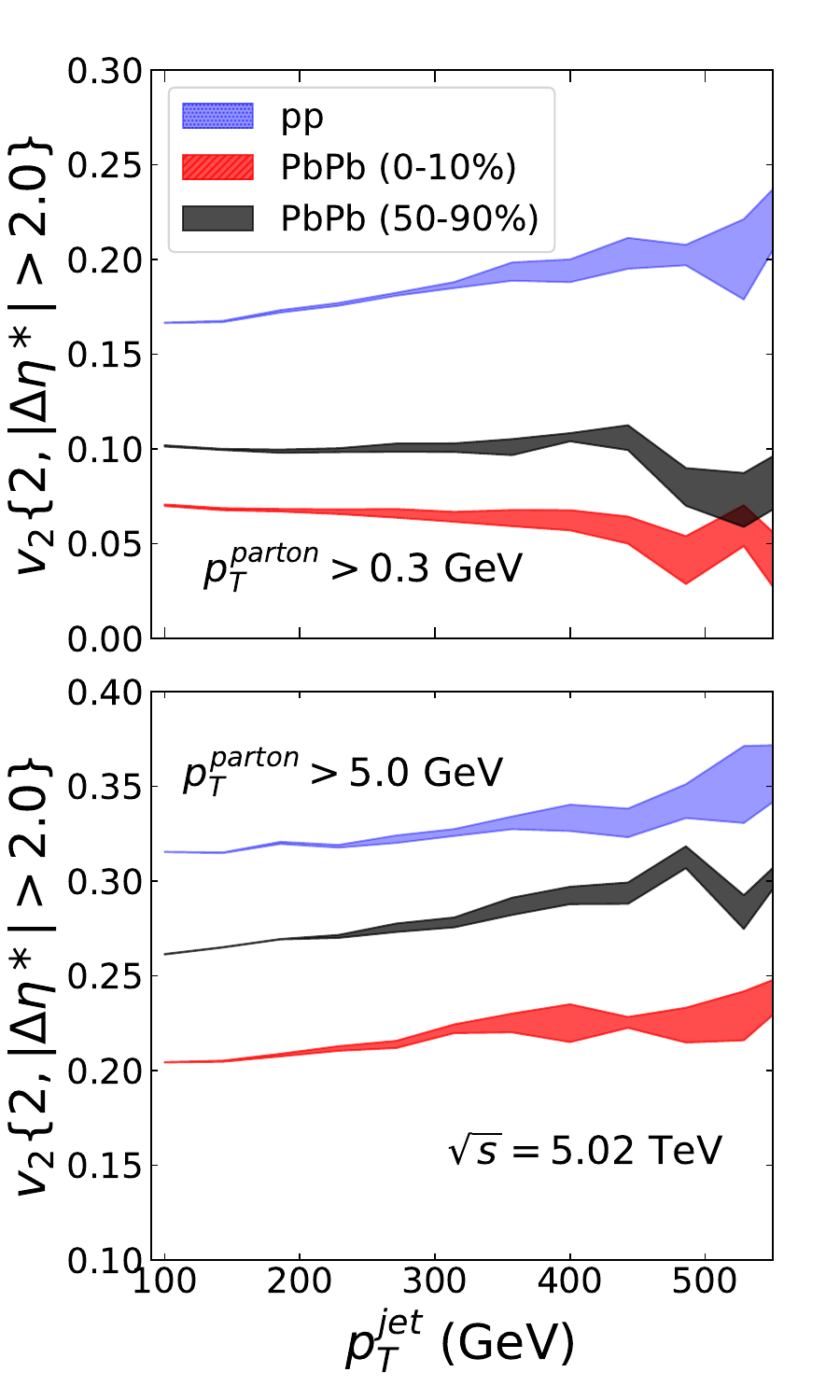}
  \caption{(Color online) Results of $v_2\{2,|\Delta \eta^*|>2.0\}$ as a function of $p_T^{\textrm{jet}}$ for inclusive jet production in $pp$ and PbPb collisions at $\sqrt{s_{NN}}=5.02$~TeV, calculated with with two cuts on transverse momenta of jet particles as $p_T>0.3$~(top panel) and $p_T>5.0$~GeV~(bottom panel), respectively. Anti-$k_T$ jets at partonic levels with $R=0.8$ selected in region $|\eta|<1.6$ are used in calculations. Results  for two centrality classes of PbPb collisions, i.e., $0\!-\!10\%$ and $50\!-\!90\%$ are shown respectively.}\label{fig:v2_pp_pb}
\end{figure}

Before the study for a more realistic nuclear medium environment in AA collisions, we show a test for a simple case with a uniform static medium. Shown in Tab.~\ref{tab:static} are the results of $v_2\{2,|\Delta \eta^*|>2.0\}$ for inclusive jet production with $p_{T}^{\textrm{jet}}>100$~GeV, where the jets are reconstructed at the partonic level and $v_2$ are calculated with $p_T^{\textrm{parton}}>0.3$~GeV. Values of the model parameters, including a medium temperature, a jet propagation time in medium, an effective strong coupling $\alpha_s$ and a type of scattering~(elastic/inelastic) suffered by the propagating partons~\cite{Luo:2018pto,He:2015pra,Cao:2016gvr,Cao:2017hhk}, are listed in the left column of Tab.~\ref{tab:static}. It is observed that, with the presence of a static medium, the values of $v_2$ are suppressed to some extent relative to the result for $pp$ collisions, and are sensitive to the model parameters. The suppression of $v_2$ is stronger either for a higher temperature, for a longer propagation time or for a higher strength of jet-medium interaction. It is also found that both the elastic and inelastic scattering processes give nontrivial contributions to the suppression of $v_2$.

It is noted that, in the calculations with $p_T^{\textrm{parton}}>0.3$~GeV, the jet particles from the medium excitations, probably carrying small fractions of the jet momenta~\cite{Luo:2018pto,Wang:2016fds,Chen:2017zte} may contribute to the modification of $v_2$. For instance, with the parameter setting 'A' in Tab.~\ref{tab:static}, $16\%$ of the jet particles are from the medium excitations in LBT model. As a test, we further calculate the $v_2$ with $p_T^{\textrm{parton}}>5$~GeV, where most of the excitation particles are excluded, and find similar suppression on $v_2$. For example, the $v_2$ calculated for $pp$ and for the medium with setting 'A' are 0.316 and 0.259, respectively. 

In order to study the more realistic cases in AA collisions, in the simulation with LBT, we use the profiles of the evolving background QGP media provided by a 3+1D CLVisc hydrodynamic calculation~\cite{Pang:2012he,Pang:2014ipa} with initial conditions from A Multi-Phase transport model~\cite{Lin:2004en}, and set the effective coupling as $\alpha_s=0.15$~\cite{He:2018xjv,Zhang:2021sua}. With similar model settings, the LBT simulations can well explain the experimental data on various jet quenching observables~\cite{Cao:2017hhk,He:2018xjv,Luo:2018pto,Zhang:2018urd,Zhang:2021sua,Chen:2020pfa}.

Figure~\ref{fig:v2_pp_pb} shows the results of $v_2\{2,|\Delta \eta^*|>2.0\}$ for inclusive jet production as a function of $p_T^{\textrm{jet}}$ in both $pp$ and PbPb collisions at $\sqrt{s_{NN}}=5.02$~TeV, where jets are reconstructed at the partonic level. It is observed that, relative to the results in $pp$ collisions, the $v_2$ in PbPb collisions are significantly suppressed with both $p_T^{\textrm{parton}}>0.3$~GeV and $p_T^{\textrm{parton}}>5$~GeV, and the suppression becomes stronger for more central collisions, in consistent with the results for static medium.

 \begin{figure}[b]
 \vspace{-15pt}
  \centering
\includegraphics[width=0.45\textwidth]{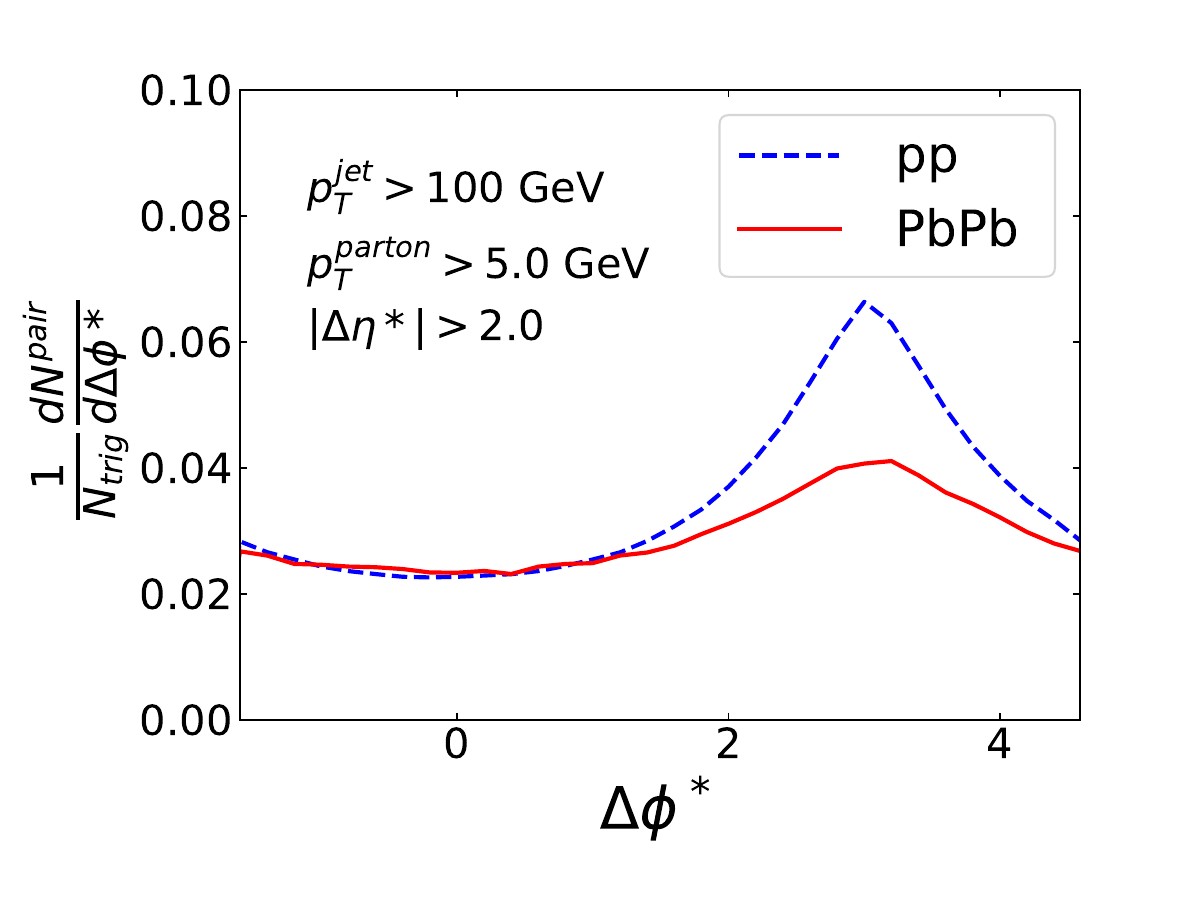} \vspace{-15pt}
  \caption{(Color online) Trigger-particle-normalized two-particle azimuthal-angle distribution inside jets as a function of $\Delta\phi^*$ for $|\Delta\eta^*|>2.0$ in both $pp$ and PbPb~($0\!-\!10\%$) collisions.}\label{fig:deltaphi}
\end{figure}

To further understand such a suppression effect, we calculate the two-particle azimuthal-angle distribution inside jets defined as $(1/{N^{trg}}){dN^{pair}}/{d\Delta\phi^*}$~\cite{CMS:2023iam}, where $N^{trg}$ and $dN^{pair}$ denote numbers of trigger particles and particle pairs, respectively, and $\Delta\phi^*$ represents the difference between the azimuthal-angles of two particles. Both the results for $pp$ and PbPb collisions calculated with partonic jets and for $p_T^{\textrm{parton}}>5$~GeV are plotted in Fig.~\ref{fig:deltaphi}. Compared to the results for $pp$ collisions, a flatter distribution is observed for PbPb, indicating a de-correlation in azimuthal angle due to the jet-medium interactions, generally related to a more isotropic configuration.
\begin{figure}[t]
    \centering
    \hspace{-0.04\textwidth} 
    \subfigure{
        \includegraphics[width=0.25\textwidth]{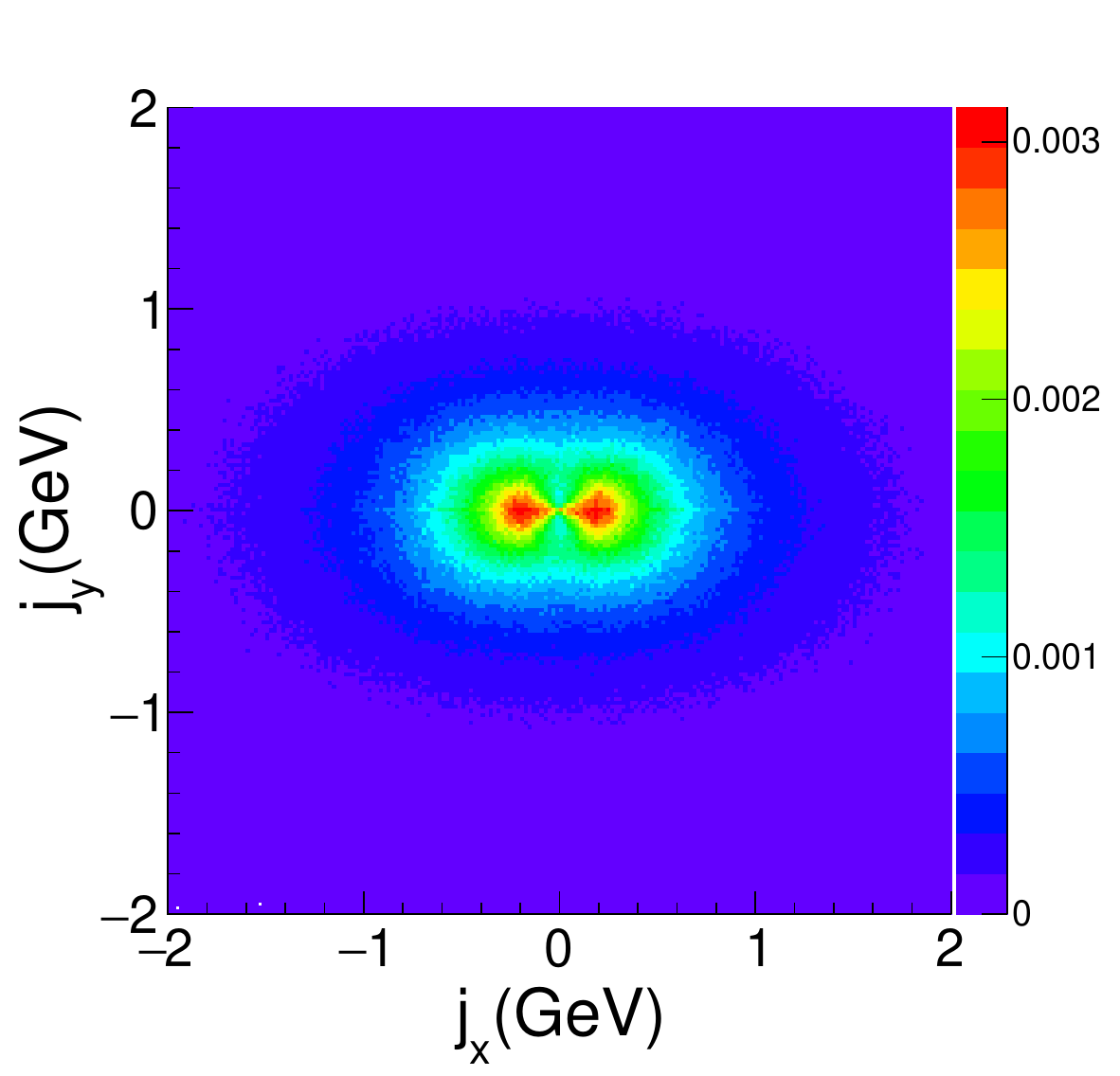}
    }
    \hspace{-0.02\textwidth} 
    \subfigure{
        \includegraphics[width=0.25\textwidth]{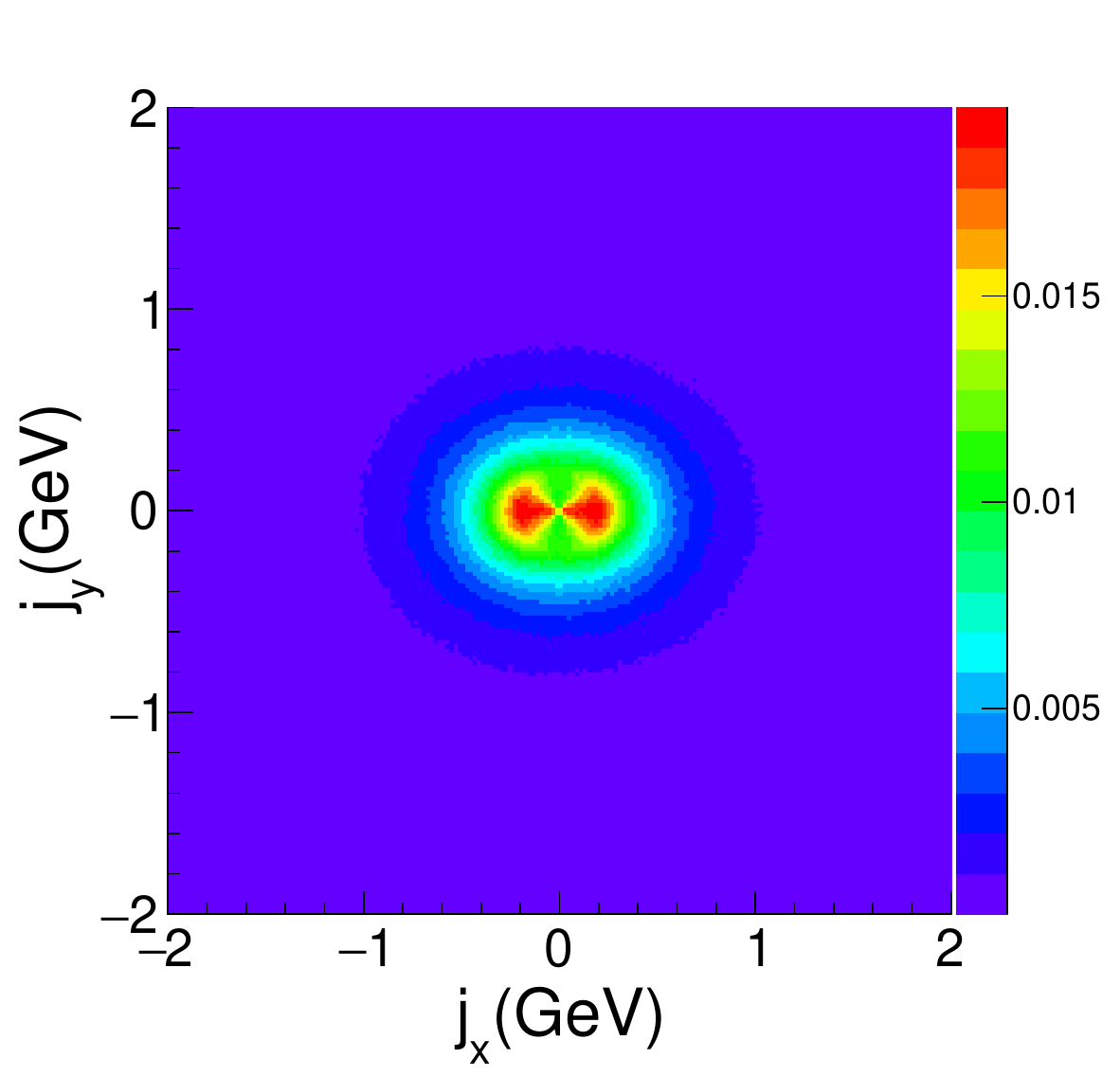}
    }
    \caption{Jet-particle distributions in $\vec{j}_T$ plane for inclusive jet production in both $pp$~(left panel) and central PbPb~(right panel) collisions at $\sqrt{s_{NN}}=5.02$~TeV. Anti-$k_T$ jets reconstructed at a partonic level with R=0.8 and $p_{T}^{jet}>100$~GeV are used. Distributions are obtained by averaging over $5\times 10^5$ collected jets with jet plane orientation in alignment.}
    \label{fig:pp_aa_jx_jy}
\end{figure}

\begin{figure}[b]
 \vspace{-15pt}
  \centering
\includegraphics[width=0.45\textwidth]{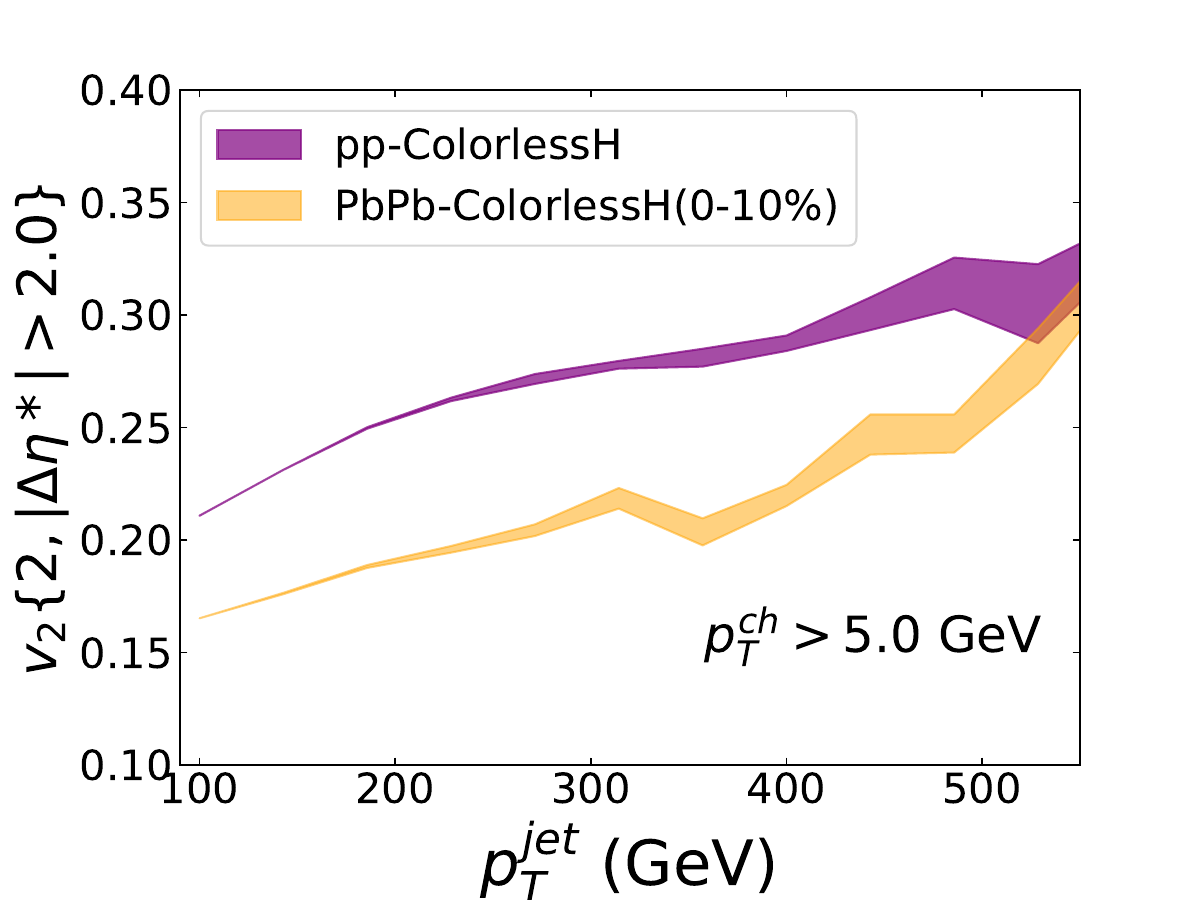}
  \caption{(Color online) Results of $v_2\{2,|\Delta \eta^*|>2.0\}$ as a function of $p_T^{\textrm{jet}}$ for inclusive jet production in $pp$ and central PbPb collisions at $\sqrt{s_{NN}}=5.02$~TeV. Jets are reconstructed at hadronic level with a colorless hadronization. Charged particles with $p_T^{\textrm{ch}}>5$~GeV are used in calculations.}\label{fig:v2_colorless}
\end{figure}

For providing a more intuitive picture of the transverse configuration inside jets and the medium modification on it, we illustrate in Fig.~\ref{fig:pp_aa_jx_jy} the particle distributions inside jets in $\vec{j}_T$ plane for inclusive jets in both $pp$ and PbPb collisions. To reserve the elliptic anisotropy in such a distribution, we  generalize the "event plane" method~\cite{Bilandzic:2010jr,Voloshin:2008dg} for the study of anisotropic flow into the jet systems, and define a second-order jet-plane angle $\Psi_{\textrm{JP},2}=\frac{1}{2}\tan^{-1}({Q_{2,y}}/{Q_{2,x}})$ with "Q-vectors" expressed as
\begin{align}
\begin{split}
    Q_{2,x}=\sum_{i}\cos(2\phi_i^*), ~~~~ Q_{2,y}=\sum_{i}\sin(2\phi_i^*)
    \end{split}
\end{align}
In the calculations for Fig.~\ref{fig:pp_aa_jx_jy}, the jet particles with $p_T^{\textrm{parton}}>0.3$~GeV are used to determine the jet plane, and for each generated jet system, the jet plane is rotated to be oriented along the $j_x$ direction~\cite{Qiu:2011hf}. By averaging over all the collected jets, one can observe in Fig.~\ref{fig:pp_aa_jx_jy} that, the jets generated in $pp$ collisions exhibit an obvious elliptic distribution, whereas the quenched jets in AA collisions becomes more isotropic. The elliptic anisotropy for such distributions can be evaluated with $v_2=\langle \frac{j_x^2-j_y^2}{j_x^2+j_y^2}\rangle$~\cite{Ollitrault:1992bk,Bilandzic:2010jr}, which are found to be 0.33 and 0.16 for $pp$ and AA, respectively. Moreover, a double-peak structure can be found in both of the two distributions, indicating that such a back-to-back configuration plays an important role in the generation of elliptic anisotropy inside jets. We observe that the values of $v_2$ calculated with such a method are larger than the results of $v_2\{2,|\Delta \eta^*|>2.0\}$ obtained with the two-particle correlation method~(Fig.~\ref{fig:v2_pp_pb}). This discrepancy can be partly attributed to the small number of particles and the less preserved longitudinal symmetry in the jet systems. Notably, despite these methodological differences, both approaches consistently predict the medium-induced suppression of $v_2$.

In the above results, the medium-induced suppression of the elliptic anisotropy inside jets is observed for jets reconstructed at a partonic level, so it will be interesting to further study the effect at a hadronic level. In general, the hadronization mechanism in AA collisions can be much more complicated than that in $pp$ collisions, and some hybrid models have been developed recently~\cite{Zhao:2020wcd,Zhao:2021vmu,JETSCAPE:2025wjn}. In Ref.~\cite{Zhao:2021vmu}, a medium freeze out, a parton coalescence and a string fragmentation are combined to simultaneously explain the data from soft to hard regions, where different regions are dominated by different mechanisms within the model~\cite{Zhao:2021vmu}. We can expect that the results in Fig.~\ref{fig:v2_pp_pb} for different cuts on jet particles may be affected by different features of hadronization. For example, the soft hadrons from the medium freeze out may exhibit a nearly thermal or isotropic behavior, whereas the hard particles may follow a fragmentation pattern, which may be studied with a comprehensive model in future.

In this work, we show a simple test with a colorless hadronization scheme~\cite{JETSCAPE:2019udz,JETSCAPE:2022jer} and plot the results of $v_2\{2,|\Delta \eta^*|>2.0\}$ for hadronic jets with $p_T^{\textrm{ch}}>5$~GeV in both $pp$ and central PbPb collisions in Fig.~\ref{fig:v2_colorless}. Suppression of $v_2$ in AA collisions can also be observed in most of the studied region. It should be mentioned that, the result for $pp$ is somewhat different from that with the string fragmentation in Fig.~\ref{fig:v2_pp_inclusive}, and a further test shows that the calculations with the same settings in PYTHIA8 but with colorless hadronization can hardly reproduce the data of $v_2\{2,|\Delta \eta^*|>2.0\}$ as a function of multiplicity in Fig.~\ref{fig:CMS}.

\section{Summary and Discussion}
\label{sec:summary}
Jet quenching has become one of the iconic phenomena demonstrating the presence of extremely hot and dense nuclear matter in high-energy nuclear collisions. As a kind of many-body probe, jets have shown wealthy new potentials to explore the properties of the QCD matters. In this work, we study the medium-induced modifications of the elliptic anisotropy inside jets for inclusive jet production in relativistic heavy-ion collisions. By simulating the jet propagation in QGP medium with an LBT model, we observe an obvious de-correlation in two-particle azimuthal angular distribution inside jets in AA collisions relative to that in $pp$ collisions, leading to significant suppression of the in-jet elliptic anisotropy coefficient $v_2$, which can be examined in future experiments.

Such an effect may be somewhat robust and universal due to the dissipative nature of the jet transport process, in which the stochastic scatterings suffered by jet particles can erase the memory of the earlier-time dynamical information inside jet. As off-equilibrium systems, the jets propagating in thermal media may become more disordered than that in vacuum. Future studies at a hadronic level with a comprehensive hadronization scheme as well as a final-state hadron cascade will be of great interest. Since the modifications on $v_2$ inside jets are found to be sensitive to the medium properties, the measurement of such observables may provide new resolution power of the jets as a microscope of the structures of QCD matters.

{\bf Acknowledgments:} The authors would like to thank E.-k.~Wang, W.-B. Zhao, H.-Z. Zhang, T. Luo, S.-L. Zhang, S. Wang, S. Shen, J. Kang, Y. Li, Y. Zhang and Q. Wang for helpful discussions. This research is supported by Guangdong Major Project of Basic and Applied Basic Research No. 2020B0301030008, by Guangdong Basic and Applied Basic Research Foundation under Project No. 2022A1515110392, and by Natural Science Foundation of China with Project Nos. 11935007 and 11805167.

\end{document}